\pdfoutput=1
\newcommand*{\ATLASLATEXPATH}{latex/}
\documentclass[cernpreprint,texlive=2016,txfonts,UKenglish]{latex/atlasdoc}
\usepackage{\ATLASLATEXPATH atlaspackage}
\usepackage{\ATLASLATEXPATH atlasbiblatex}

\usepackage{\ATLASLATEXPATH atlascontribute}

\usepackage{\ATLASLATEXPATH atlasphysics}
\usepackage{pdflscape}
\usepackage{longtable}
\usepackage{multirow}

\graphicspath{{logos/}{figures/}}

\usepackage{latex/atlasmetdefs}

\newcommand{\fixme}[1]{}
\newcommand{\cutoff}[1]{}
\newcommand*{\lumi}{\mbox{$\;$(lumi.)}\xspace}

\DeclareOldFontCommand{\rm}{\normalfont\rmfamily}{\mathrm}
\DeclareOldFontCommand{\tt}{\normalfont\ttfamily}{}
\DeclareOldFontCommand{\it}{\normalfont\itshape}{}

\usepackage{dcolumn}
\newcolumntype{U}{D{,}{\,\pm\,}{-1}}
\newcolumntype{P}{D{,}{.}{-1}}

\def\kt{\ensuremath{k_t}}
\def\antikt{anti-\kt}
\def\aznlo{\texttt{AZNLO}}
\def\ctsix{\texttt{CTEQ6L1}}

\def\cttennlo{\texttt{CT10 NLO}}
\def\nnpdf{\texttt{NNPDF3.0}}
\def\mstw{\texttt{MSTW2008nlo}}
\def\nnpdflo{\texttt{NNPDF2.3 LO}}
\def\sherpa{\texttt{SHERPA}} 
\def\evtgen{\texttt{EvtGen}}

\def\herwigpp{\texttt{HERWIG++}}

\def\amcatnlo{\texttt{Madgraph5\_aMC@NLO}}

\def\openloops{\texttt{OpenLoops}} 

\def\powhegbox{\texttt{POWHEG-BOX}}
\def\pythia{\texttt{PYTHIA}}
\def\pythiasix{\pythia\ {\tt v6.428}}
\def\photos{\texttt{PHOTOS++}}
\def\pythiaeight{\pythia\ {\tt v8.210}}

\def\openloops{\texttt{OpenLoops}}

\def\qqtoWW{$q\bar{q} \to \Wboson\Wboson$\xspace}
\def\ggtoWW{$gg \to \Wboson\Wboson$\xspace}
\def\ggtoHtoWW{$gg \to H \to \Wboson\Wboson$\xspace}
\def\WW{\ensuremath{WW}\xspace}
\def\WWemu{\ensuremath{WW \to e \mu}\xspace}
\def\metrel{\ensuremath{E_{\mathrm{T,~Rel}}^{\mathrm{miss}}}\xspace}
\def\ptmiss{\ensuremath{p_{\mathrm{T}}^{\mathrm{miss}}}\xspace}

\AtlasTitle{Measurement of the $W^+W^-$ production cross section in $pp$ collisions at a centre-of-mass energy of $\sqrt{s}=13$ \TeV\ with the ATLAS experiment}

\author{The ATLAS Collaboration}

\PreprintIdNumber{CERN-EP-2016-267}

\AtlasDate{\today}

\AtlasJournal{Phys.\ Lett.\ B.\ }
\AtlasJournalRef{\PLB ~773 (2017) 354}
\AtlasDOI{10.1016/j.physletb.2017.08.047}

\AtlasAbstract{%
The production of opposite-charge \Wboson-boson pairs in proton--proton collisions at $\sqrt s = 13$~\TeV\ is measured using data corresponding to 3.16 \ifb\ of integrated luminosity collected by the ATLAS detector at the CERN Large Hadron Collider in 2015. Candidate \Wboson-boson pairs are selected by identifying their leptonic decays into an electron, a muon and neutrinos. Events with reconstructed jets are not included in the candidate event sample. The cross-section measurement is performed in a fiducial phase space close to the experimental acceptance and is compared to theoretical predictions. Agreement is found between the measurement and the most accurate calculations available.
}

\hypersetup{pdftitle={ATLAS WW Run 2},pdfauthor={The ATLAS Collaboration}}

\begin{document}

\maketitle

\section{Introduction}
\label{sec:intro}

The measurement of the production properties of opposite-charge \Wboson-boson pairs (denoted by \WW in this Letter) is an important test of the Standard Model (SM) of particle physics. This process is sensitive to the strong interaction between quarks and gluons and probes the electroweak gauge structure of the SM.  

Measurements of \WW production were first conducted at LEP~\cite{schael:2013ita} using electron--positron collisions. Measurements in hadron collisions were first carried out at the Tevatron by the CDF~\cite{Abe:1996dw,Aaltonen:2009aa} and D\O{}~\cite{Abazov:2009ys} Collaborations. At the Large Hadron Collider (LHC), the \WW production cross sections have been measured in proton--proton collisions for centre-of-mass energies of $\sqrt{s}=7$~\TeV\ and $\sqrt{s}=8$~\TeV\ by the ATLAS \cite{ww7tev,STDM-2013-07} and CMS~\cite{CMS-SMP-12-005,Khachatryan:2015sga} Collaborations. 
In order to match the experimental precision and address discrepancies between data and theory reported in some of the 8 \TeV\ results, significant progress has been made in theoretical calculations to include higher-order corrections in perturbative Quantum Chromodynamics (pQCD) ~\cite{Gehrmann:2014fva,Meade:2014fca,Jaiswal:2014yba,Monni:2014zra,Grazzini:2016ctr,Dawson:2016ysj}.  
The \WW signal is composed of three leading sub-processes: \qqtoWW production\footnote{In this Letter, the notation \qqtoWW is used to include both the $q\bar{q}$ and $qg$ initial states for $\Wboson\Wboson$ production.} (in the $t$- and $s$-channels\fixme{ u channel too?}), non-resonant \ggtoWW production, and resonant \ggtoHtoWW production (with both $gg$-initiated processes occurring through a quark loop).
These sub-processes are known theoretically at different orders in the strong coupling constant $\alpha_{\mathrm{s}}$. 

This Letter describes a measurement of \WW production in proton--proton collisions at $\sqrt{s}=13$~\TeV\ with the ATLAS detector using the data collected during the 2015 run. The cross section is measured within a phase space close to the geometric and kinematic acceptance of the experimental analysis, i.e.~a fiducial phase space, in the  $WW \rightarrow e^\pm \nu \mu^\mp \nu$ (denoted in the following by \WWemu) decay channel. 
In addition, the ratio of cross sections at 13~\TeV\ and 8 \TeV\ centre-of-mass energies in the respective fiducial phase spaces is presented. 
Both measurements are compared to the latest theoretical predictions.


\section{The ATLAS detector}
\label{sec:detector}

The ATLAS detector~\cite{detectorpaper,ATLAS-TDR-19} is a multi-purpose particle detector with a cylindrical geometry.\footnote{The ATLAS experiment uses a right-handed coordinate system with its origin at the nominal $pp$ interaction point at the centre of the detector. The positive $x$-axis is defined by the direction from the interaction point towards the centre of the LHC ring, with the positive $y$-axis pointing upwards, while the beam direction is along the $z$-axis. Cylindrical coordinates ($r$, $\phi$ ) are used in the transverse ($x$, $y$) plane, $\phi$ being the azimuthal angle around the beam direction. The pseudorapidity is defined in terms of the polar angle $\theta$ from the $z$-axis as $\eta = - \ln [\tan(\theta/2)]$. The distance in $\eta$--$\phi$  space between two objects is defined as $\Delta R \equiv \sqrt{({\Delta\eta})^2+({\Delta\phi})^2}$.
Transverse energy is computed as $\et = E \cdot \sin\theta$.}
 It consists of layers of inner tracking detectors surrounded by a superconducting solenoid, calorimeters, and a muon spectrometer. The inner detector (ID) is situated inside a 2 T magnetic field generated by the solenoid and provides precision tracking for charged particles with pseudorapidity $|\eta| <$ 2.5. The calorimeter covers the pseudorapidity range $|\eta| <$ 4.9. Within $|\eta| <$ 2.47 the finely segmented electromagnetic calorimeter identifies electromagnetic showers and measures their energy and position, providing electron identification together with the ID. The muon spectrometer (MS) surrounds the calorimeters and includes three large air-core toroidal superconducting magnets with eight coils each, providing muon identification and measurement in the region $|\eta| <$ 2.7 and triggering in the region $|\eta| <$ 2.4.
A two-level trigger system is used to select events in real time. It consists of a hardware-based first-level trigger and a software-based high-level trigger. The latter uses reconstruction software with algorithms similar to the offline versions.


\section{Data and Monte Carlo samples}
\label{sec:sample}

The analysis is based on data collected with the ATLAS detector during the 2015 data-taking period. Events with $pp$ collisions at $\sqrt{s}$ = 13 \TeV\ and all relevant detector components functional have been used. This data sample corresponds to an integrated luminosity of $\cal{L}$ = 3.16 fb$^{-1}$.
  
Monte Carlo (MC) event generators are used to model signal and background processes. The $WW$, $WZ$, and $ZZ$ diboson processes (where $Z$ stands for $Z/\gamma^{*}$) with $q\bar{q}$ initial states are simulated at next-to-leading order (NLO) in pQCD with the \powhegbox\ {\tt v2} event generator~\cite{Nason:2004rx, Frixione:2007vw, Alioli:2010xd, Melia:2011tj, Nason:2013ydw} using the \cttennlo~\cite{CTEQ10} parton distribution functions (PDFs). 
For the modelling of the parton shower and non-perturbative effects such as fragmentation and the underlying event, \powhegbox\ {\tt v2} is interfaced to \pythia\ {\tt v8.210}~\cite{Sjostrand:2014zea} with the \aznlo~\cite{AZNLO:2014} set of tuned parameters and the \ctsix ~\cite{Pumplin:2002vw} PDF. The invariant mass of the leptons originating from the $Z$ boson or photon in the $ZZ$ and $WZ$ samples is required to satisfy $m_{\ell\ell} > 7$ \GeV. 
A sample of $WZ$ events generated with \sherpa\ {\tt v2.1.1}~\cite{Sherpa} with $m_{\ell\ell} >$ 0.45~\GeV~is used to study systematic uncertainties.
The cross sections given by the event generator are at NLO in QCD while the $WW$, $WZ$ and $ZZ$ samples are normalised using their respective inclusive next-to-next-to-leading order (NNLO) predicted cross sections~\cite{Gehrmann:2014fva,Grazzini:2016swo,Grazzini:2015hta,Cascioli:2014yka}.
The configuration of the \powhegbox\ {\tt v2} event generator, as described above, reproduces the distribution predicted by NNLO calculations matched to resummation calculations up to next-to-next-to-leading logarithm (NNLL)~\cite{Gehrmann:2014fva,Meade:2014fca} for the transverse momentum of the $WW$ system ($p_{\mathrm T}^{WW}$)  in the range relevant to this analysis, so no further steps are taken to explicitly include resummation effects in the $WW$ signal samples.
The resonant \ggtoHtoWW signal contribution is simulated
with the \powhegbox\ {\tt v2} event generator~\cite{Bagnaschi:2011tu} and
normalised using the inclusive next-to-next-to-next-to-leading order (N$^3$LO) predicted cross section~\cite{Anastasiou:2015ema}. 
The non-resonant \ggtoWW
signal contribution is modelled with \sherpa\ {\tt v2.1.1} at leading order (LO) using
\openloops\ with up to one additional parton in the final
state~\cite{jetvetoNNLO} and normalised using the inclusive NLO
predicted cross section~\cite{Caola:2015rqy}.

The $Z$($\to ee/\mu\mu/\tau\tau$)+jets production processes are simulated with the
\amcatnlo\ {\tt v2.2.2} \cite{Alwall:2014hca} event generator interfaced to {\tt Pythia v8.186}. The matrix elements for $Z$ production with up to four associated partons are calculated at LO and the \nnpdflo\ PDF set~\cite{Ball:2012cx} is used. The \photos\ program version {\tt 3.52} ~\cite{Golonka:2005pn} is used for QED emissions from electroweak vertices and charged leptons. Alternative samples of $Z$($\to \tau\tau$)+jets are produced with different MC event generators for the estimation of systematic uncertainties in the modelling: \powhegbox\ {\tt v2} with NLO matrix elements interfaced to \pythiaeight, and \sherpa\ {\tt v2.2.0} with NLO matrix-element accuracy up to two associated partons and with LO accuracy for three and four associated partons. The $Z$+jets events are normalised using the NNLO $Z$ production cross section~\cite{Anastasiou:2003ds}.

The \sherpa\ {\tt v2.1.1} event generator is used to model the $W\gamma$ and $Z\gamma$ processes with LO matrix element calculations for events with up to 3 partons in the final state matched to parton shower, using the \cttennlo\ PDF set and with the $\gamma$ transverse momentum greater than 10~\GeV.

The \powhegbox\ {\tt v2} event generator~\cite{Campbell:2014kua,Re:2010bp} with the \cttennlo\ PDF is used for the generation of $t\bar{t}$ and single top quarks in the $Wt$ channel. Parton shower, fragmentation, and the underlying event are simulated using
\pythia\ {\tt v6.428}~\cite{Sjostrand:2006za} with the \ctsix\ PDF
and the {\tt Perugia 2012}~\cite{Skands:2010ak} set of parameters. The top-quark
mass is set to 172.5 \GeV. 
Alternative samples are generated with different settings to assess the uncertainty in modelling top-quark events. For estimating the effect of parton shower and hadronisation modelling an alternative sample is generated with the \powhegbox\ {\tt v2} event generator interfaced to \herwigpp~\cite{Gieseke:2011na}. 
A comparison between this sample and a different one produced with \amcatnlo\ interfaced to \herwigpp\ is used to estimate the uncertainty associated to the matrix-element implementation and the matching to the parton showers. Separate alternative samples are also generated with \powhegbox\ {\tt v2} interfaced to \pythiasix\ with extra jet radiation emitted in the matrix element and in the parton shower. 
In addition, the modelling of the overlap at NLO between $Wt$ and $t\bar{t}$ diagrams~\cite{Frixione:2008yi} is studied. The effect is assessed by generating $Wt$ events with different schemes for overlap removal using the \powhegbox\ {\tt v2} event generator interfaced to \pythiasix\ for the simulation of parton showering and non-perturbative effects.
These samples are simulated following the recommendations documented in Ref.~\cite{topMCs}.
The $t\bar{t}$ samples are normalised using the NNLO+NNLL soft-gluon resummation
prediction~\cite{Czakon:2011xx}, while the $Wt$ samples are normalised using the NLO+NNLL prediction~\cite{Kidonakis:2010ux}.

The \evtgen\ {\tt v1.2.0}~\cite{evtgen} program is used for the properties of the bottom and charm hadron decays in all samples generated using the \powhegbox\ {\tt v2} and \amcatnlo\ {\tt v2.2.2} programs. The generated samples are passed through a simulation of the ATLAS detector based on GEANT4~\cite{Geant4,Aad:2010ah}. 
They are overlaid with additional proton--proton interactions (pile-up) generated with \pythiaeight\ and the distribution of the average number of interactions per bunch crossing is reweighted
to agree with the corresponding data distribution. The
simulated events are reconstructed and analysed with
the same algorithms as the data and are corrected with data-driven correction factors to account for differences between data and simulation in lepton and jet reconstruction and identification.


\section{Event reconstruction and selection}
\label{sec:selection}
The \WW event candidates are selected by requiring exactly one electron and one muon of opposite charge in the event, and significant missing transverse momentum, as described below. Events with a same-flavour lepton pair are not used because they have larger background from the Drell--Yan process. 

Candidate events are preselected by either a single-muon or single-electron trigger requiring transverse momentum \pt > 20 or 24 \GeV\ respectively. The efficiency of the trigger for selecting $WW$ events is approximately $99\%$ for events that pass the offline selection.

Leptons are required to originate from the primary vertex, defined as the reconstructed vertex with the largest sum of the $\pt^2$ of the associated tracks. The longitudinal impact parameter of each lepton track, defined as the distance along the beam line between the track and the point of closest approach of the track to the primary vertex, multiplied by the sine of the track $\theta$ angle, is required to be less than 0.5 mm. Furthermore, the significance of the transverse impact parameter calculated with respect to the beam line, $|d_0/\sigma_{d_0}|$, is required to be less than 3.0 (5.0) for muons (electrons). 

Electron candidates are reconstructed from the combination of a cluster of energy deposits in the electromagnetic calorimeter and a track in the ID~\cite{ATLAS-CONF-2016-024}. Candidate electrons must satisfy the {\it Tight} quality definition described in Ref.~\cite{ATLAS-CONF-2016-024}. Muon candidates are reconstructed by combining a track in the ID with a track in the MS~\cite{Aad:2016jkr}. The {\it Medium} criterion, as defined in Ref.~\cite{Aad:2016jkr}, is applied to the combined tracks. The leptons are required to be isolated using information from ID tracks and calorimeter energy clusters in a cone around the lepton. The expected isolation efficiency for prompt leptons is at least 90\% (99\%) at a \pt of 25 (60)~\GeV~using a so-called {\it gradient} working point~\cite{Aad:2016jkr,ATLAS-CONF-2016-024}.  

Jet candidates are reconstructed within the calorimeter acceptance using the anti-\kt~jet clustering algorithm \cite{antikt-jet} with a radius parameter of $R$ = 0.4 which combines clusters of topologically-connected calorimeter cells \cite{ATL-PHYS-PUB-2015-036}.  The jet energy is calibrated by applying a \pt- and $\eta$-dependent correction derived from MC simulation with additional corrections based on data \cite{ATL-PHYS-PUB-2015-015}. As part of the jet energy calibration a pile-up correction based on the concept of jet area is applied to the jet candidates \cite{ATLAS-CONF-2013-083}. 
The jet-vertex-tagger (JVT) technique \cite{ATLAS-CONF-2014-018} is used to separate hard-scatter jets from pile-up jets within the acceptance of the tracking detector by requiring a significant fraction of the jet's summed track \pt to come from tracks associated with the primary vertex. 
A jet-vertex-tagger requirement of JVT $> 0.64$ for jets with $\pt<50$ \GeV\ and $|\eta|<2.4$ is applied. This requirement has an efficiency that increases with the jet \pt and is between $87\%$ and $98\%$ for selecting hard-scatter jets with \pt in the range 20--50 \GeV. 
Candidate jets are discarded if they lie within a cone of size $\Delta R = 0.2$ around an electron or, for jets with less than three associated tracks, around a muon candidate. If a jet with three or more associated tracks lies within $\Delta R < 0.4$ of a muon, or $0.2 < \Delta R < 0.4$ of an electron, the corresponding lepton candidate is discarded.
Within the ID acceptance, jets originating from the fragmentation of $b$-hadrons ($b$-jets) are identified using a multivariate algorithm \cite{ATL-PHYS-PUB-2015-022, ATL-PHYS-PUB-2015-039}. The chosen operating point has an efficiency of 85\% for selecting jets containing $b$-hadrons and a rejection factor of 28 for light-quark jets, as estimated in a sample of simulated $t\bar{t}$ events and validated with data.

The missing transverse momentum is computed as the negative of the vectorial sum of the transverse momenta of the reconstructed objects selected in the analysis (i.e.\ electrons, muons, and jets), and a soft term based on the tracks associated with the primary vertex but not with the hard objects explicitly used in the missing transverse momentum computation \cite{ATL-PHYS-PUB-2015-023}. The magnitude of the missing transverse momentum is denoted by \met in the following.
The jet selection in the \met computation is chosen to provide a compromise between good resolution and scale, with the requirement of \pT> 20~\GeV\ for all jets, and an additional JVT $>$ 0.64 requirement for jets in the region of $|\eta|$ $<$ 2.4. \cutoff{\met is computed as the magnitude of the vector sum of the transverse momenta of hard, identified and calibrated objects (electrons, muons, jets and photons) and a soft term based on the tracks associated with the primary vertex but not with the hard objects explicitly used in the \met computation \cite{ATL-PHYS-PUB-2015-023}.  Since low \pt jets are sensitive to pileup effects, the choice of jet selection influences the balance between the hard objects and the soft terms, and is optimized for this analysis.}In Drell--Yan production of $\tau$-lepton pairs with subsequent decay to an $e$--$\mu$ pair, the direction of the missing transverse momentum tends to align with a final-state lepton. To suppress this contamination a requirement is imposed on the missing transverse momentum component perpendicular to the direction in the $r$--$\phi$ plane of the lepton closest to the missing transverse momentum direction, as defined in Ref.~\cite{STDM-2013-07}. This variable is denoted in the following by \metrel. 
In addition, a more pile-up-robust track-based missing transverse momentum variable of magnitude \ptmiss is computed within the ID acceptance~\cite{ATL-PHYS-PUB-2015-023}, using only ID tracks associated with the primary vertex.

The signal region (SR) in which the measurement is performed is defined as follows. Candidate $WW$ events are required to have one electron and one muon, each with \pt $>$ 25~\GeV, of opposite charge. The electron is required to be in the region $|\eta| <$ 2.47, excluding the transition region between the barrel and endcap calorimeters. For the muon, $|\eta| <$ 2.4 is required. To reduce the background from other diboson processes,  the events are required to have no additional electron or muon with \pt $>$ 10~\GeV. To suppress the background contribution from top quarks, events are required to have no jets with \pt$>$ 25 (30)~\GeV\ in $|\eta|<$ 2.5 (4.5), and no $b$-jets with \pt $>$ 20~\GeV. In addition, the requirements \metrel $>$ 15~\GeV,  \ptmiss $>$ 20~\GeV, and the invariant mass of the lepton pair $m_{e\mu} >$ 10~\GeV\ suppress Drell--Yan background contributions. 
The lepton, jet, and event selection criteria are summarised in Table~\ref{tab:recolevelcuts}. 

\begin{table}[!ht]\renewcommand{\arraystretch}{1.2}
  \begin{center}
  \begin{tabular}{l|c}
    \hline\hline
Selection requirement                            &   Selection value \\
\hline\hline
$ p_{\rm T}^\ell$     	&   {$>25$~\GeV} 		\\
\hline
$ \eta^{\ell}$     				&  $|\eta^{e}|<2.47$ (excluding $1.37<|\eta^{e}|<1.52$), \\
   &   $|\eta^{\mu}|<2.4$  \\
\hline
Lepton identification				&  Tight (electron), Medium (muon)                                                    \\
\hline
Lepton isolation				&  Gradient working point                \\
\hline
Number of additional leptons ($\pt>10$ \GeV)     	&   0					\\
\hline
$ m_{e\mu}$     				&   $>10$ \GeV 			\\
\hline
Number of jets with $\pt>$25(30) \GeV, $|\eta|<$2.5(4.5)& 0					\\
\hline
Number of $b$-tagged jets ( $\pt>20$ \GeV, 85\% op. point) 	& 0 					\\
\hline
\metrel    					&   $>15$ \GeV 	\\
\hline
\ptmiss    					&   $>20$ \GeV 			\\

\hline\hline
  \end{tabular}
  \end{center}
  \caption{Lepton, jet, and event selection criteria for $WW$ candidate events. In the table $\ell$ stands for $e$ or $\mu$. 
  The definitions of identification and isolation are given in Refs. \cite{ATLAS-CONF-2016-024} and \cite{Aad:2016jkr}. 
 }
  \label{tab:recolevelcuts}
\end{table}


\section{Background estimation}
\label{sec:backgrounds}
 After applying the event selection requirements described in Section~\ref{sec:selection}, the dominant background in the \WW candidate sample is top-quark ($t \bar{t}$ and single top) production with neither jet nor $b$-jet above the veto thresholds within the acceptance.
Drell--Yan production of a $\tau$-lepton pair that decays leptonically can also give rise to the $e\mu$ final state. Multi-jet production with two jets misidentified as leptons, or $W+$jets production with leptonic $W$ decay and a jet misidentified as a lepton (collectively referred to as $W+$jets background below) can be mistakenly accepted as candidate events. 
This background category includes events where an electron or a muon is produced from a semileptonic decay of a bottom or charm hadron and $WW$ events where one $W$ decays leptonically and the other hadronically.
Other diboson ($WZ$, $ZZ$, $W\gamma$ and $Z\gamma$) production contributes a smaller background. Minor background processes are modelled with MC simulations, while data-driven methods are used to determine the dominant backgrounds and backgrounds with a misidentified lepton.
The normalisations of top-quark and Drell--Yan backgrounds are determined from dedicated control regions after a simultaneous fit, described in detail in Section~\ref{sec:result}. 
The phase spaces of top-quark and Drell--Yan control regions are chosen to be close to the one of the signal region.
Modelling uncertainties for each of the backgrounds, discussed here, as well as the systematic and statistical uncertainties given in Section~\ref{sec:result}, are included as nuisance parameters in the fit.

The top-quark background control region is defined by requiring one jet with \pt > 25 \GeV\ and at least one $b$-jet with \pt > 20 \GeV\ in the ID acceptance region of $|\eta|<2.5$, in an event sample selected with the same lepton criteria as the signal region and no requirement on \metrel. This control region has an estimated top purity of 93\%.  
The top-quark background, comprising $t\bar{t}$ and $Wt$ contributions, is normalised to data in this control region and both the detector and modelling uncertainties affect the extrapolation of $t\bar{t}$ and $Wt$ from the control region to the signal region.
These include $t\bar{t}$ ($Wt$) cross section uncertainties of $6\%$ ($10\%$) as well as the modelling of the parton shower and initial-state jet radiation. For the $t\bar{t}$ process the uncertainties also include the choice of MC matrix-element generator, while for the $Wt$ process they include the modelling of the overlap and interference at NLO between $Wt$ and $t\bar{t}$ diagrams estimated by comparing the nominal $Wt$ MC sample with an alternative sample generated with a different scheme for overlap removal. The uncertainties in modelling the $t\bar{t}$ and $Wt$ processes are estimated by comparing the results from the different MC samples presented in Section~\ref{sec:sample}.

The event characteristics of $e\mu$ final states from Drell--Yan production of $\tau$-lepton pairs include an $e\mu$ invariant mass below the $Z$ mass, and lower \met. 
In the Drell--Yan background control region, the $e\mu$ invariant mass is required to be $45<m_{e\mu}<80$~\GeV, and either or both of the \metrel and \ptmiss requirements are reversed to make the sample orthogonal to that in the signal region while all other selection requirements remain the same. 
The Drell--Yan control region has a purity of about $95\%$, and the Drell--Yan modelling uncertainties are taken into account by comparing different MC event generators, as discussed in Section~\ref{sec:sample}. 

Determining the background from $W$+jets production requires good knowledge of the lepton misidentification rate, which is best derived from data. 
The yield from $W$+jets production is estimated using data event samples that are selected with different lepton selection criteria: a {\it loose} lepton identification criterion is defined, leptons are selected using either the loose or the default (as used in the signal region) lepton identification criteria, and events are classified according to whether the leptons, that all satisfy the loose criteria, satisfy or not the default identification criteria.
With the introduction of the efficiencies of the default lepton identification relative to the loose lepton identification for both real and misidentified leptons, a system of four equations can be solved to estimate the number of events meeting the default lepton identification criteria.  This follows the same procedure as that described in Ref.~\cite{STDM-2013-07}.  For electrons, the loose identification corresponds to the {\it medium} criterion defined in Ref.~\cite{ATLAS-CONF-2016-024} without isolation requirements. For muons, the loose identification is the same as the default one, except that the isolation requirement is omitted.  The efficiencies for jet misidentification are determined for electrons and muons separately as a function of the lepton \pt and are cross-checked with a two-dimensional parameterisation in the lepton \pt and $\eta$. 
These efficiencies are measured using data in a control region with one lepton, at least one jet and requirements on the lepton-\met transverse mass and \met to suppress the prompt-lepton contribution from $W$+jet production. The remaining $W$+jets contribution in the selected control sample is subtracted using the MC prediction.
The efficiencies for real leptons are determined from $WW$ MC simulations with correction factors obtained by comparing $Z\rightarrow\ell\ell$ events in data and MC simulation.  
The systematic uncertainties for lepton misidentification include variations of the control region definition, the cross section uncertainties used for the subtraction of the contributions from real leptons, and the method bias (non-closure), which is estimated by comparing the prediction for the $W$+jet background contribution from MC simulation with the result of the experimental method applied to the same MC sample. 

The estimate of the diboson background from $WZ$, $ZZ$, $W\gamma$ and $Z\gamma$ processes is based on MC simulation.
The diboson background uncertainty is estimated by comparing the yields of the dominant process, $WZ$, predicted by two different event generators, \sherpa\ and \powhegbox, for which a difference of 30\% is observed. Such uncertainty is then applied to the whole diboson contribution. 

The observed numbers of events in the signal region, and the top-quark and Drell--Yan control regions, are shown later in Table~\ref{tab:Postfit_EventsTable}.


\section{Fiducial cross-section definition}
\label{sec:xsectdef}

The \WW cross section is evaluated in the fiducial phase space of the e$\mu$ decay channel. The fiducial phase space is defined in Table \ref{tab:fiducialtruthlevelcuts} as selection criteria for MC events with no detector simulation.
Electrons and muons are required at particle level to stem from one of the $W$ bosons produced in the hard scatter and their respective momenta after QED final-state radiation are vectorially added to the momenta of photons emitted in a cone of size $\Delta R = $ 0.1 around the lepton direction.
Final-state particles with lifetime greater than 30 ps are clustered into jets (referred to as particle-level jets) using the same algorithm as for detector-level jets, i.e.\ with the anti-$k_t$ algorithm with radius parameter $R$ = 0.4.
The selected charged leptons and neutrinos from \Wboson-boson decays are not included in the jet clustering.
The fiducial phase space at particle level does not make any requirement on $b$-quark jets. 
The missing transverse momentum is defined at particle level as the transverse component of the vectorial sum of the neutrino momenta. In Table \ref{tab:fiducialtruthlevelcuts}, the missing transverse momentum magnitude is denoted as $\met$, while its component perpendicular to the closest lepton in the $r$--$\phi$ plane is denoted as $\metrel$.

The fiducial cross section is defined as 
\begin{equation}
  \sigma^{\mathrm{fid}} _{\WWemu}
  = \frac{N_{\rm obs} - N_{\rm bkg}}{C\, \times {\cal L} },
\label{xSectFid}  
\end{equation}
where ${\cal L}$ is the integrated luminosity, $N_{\rm obs}$ is the observed number of events, $N_{\rm bkg}$ is the estimated number of background
events and $C$ is a factor that accounts for detector inefficiencies and contributions from $\tau$-lepton decays. 
The factor $C$ is estimated in simulation as the ratio of the number of signal events with one electron and one muon (including those from $\tau$ decays) passing the selection requirements at detector level listed in Section~\ref{sec:selection} to those passing the fiducial selection (excluding  $W\rightarrow\tau\nu$ decays) at particle level.
Therefore $C$ implicitly corrects for the contribution of $W\rightarrow\tau\nu$ decays, which is estimated in MC simulations to be $8\%$, based on their acceptance relative to the signal \WWemu channel and the relative branching fractions from the MC simulation.

\begin{table}[!htbp]
\begin{center}
\begin{tabular}{c|c}
\hline\hline
 Fiducial selection requirement  & Cut value \\
\hline\hline
 $\pt^{\ell}$    &       $>25$~\GeV  \\
$ |\eta_{\ell}|$                                        &       $<2.5$                  \\
$ m_{e\mu}$                             &       $>10$~\GeV               \\
Number of jets with $p_T>$ 25(30) \GeV, $|\eta| <$ 2.5(4.5) & 0 \\
$\metrel$                                               &       $>15$~\GeV \\
$\met$                                          &       $>20$~\GeV \\
\hline\hline
\end{tabular}
 \end{center}
\caption{Definition of the $WW\rightarrow e\mu$ fiducial phase space, where $\ell=e$ or $\mu$. }
\label{tab:fiducialtruthlevelcuts}
\end{table}


\section{Systematic uncertainties}
\label{sec:systematics}
Systematic uncertainties in the $WW$ cross-section measurement in the fiducial phase space arise from the reconstruction of leptons and jets, the background determination, pile-up and luminosity uncertainties, and the procedures used to correct for detector effects.

The uncertainty in the $C$ factor in Eq.~(\ref{xSectFid}) is dominated by experimental sources. Uncertainties in the lepton and jet reconstruction affect the signal acceptance in the fiducial phase space. The effects are estimated by varying the energy or momentum scale and the resolution of leptons and jets, and the correction factors for the trigger, reconstruction, identification and isolation efficiencies, within their uncertainties estimated in dedicated data analyses~\cite{ATLAS-CONF-2016-024,Aad:2016jkr,ATL-PHYS-PUB-2015-015}. 
Uncertainties in the \met reconstruction and $b$-tagging are also taken into account based on the studies in Ref.~\cite{ATL-PHYS-PUB-2015-023} and Ref.~\cite{ATL-PHYS-PUB-2015-022} respectively. 
The impact of the hard-object uncertainties in the \met is estimated by individually varying each of their associated uncertainties and recalculating \met for each variation. In addition, uncertainties in the scale and resolution of the \met soft term are estimated using data as discussed in Refs.~\cite{ATL-PHYS-PUB-2015-023} and~\cite{Aad:2016nrq}. 

The full set of detector-related uncertainties is taken into account in the background estimation. The statistical uncertainties stemming from the size of the MC samples used for the background estimates and from the size of the data samples used for data-driven estimations in the control regions are also considered as systematic uncertainties.  
The uncertainties due to the modelling of background processes in the signal and control regions are estimated by comparing different event generators, as discussed in Sections~\ref{sec:sample} and~\ref{sec:backgrounds}.

The MC samples are reweighted to reproduce the distributions in data of the average number of interactions per bunch crossing, and additionally the number of reconstructed primary vertices per event. The uncertainty due to pile-up is estimated as the difference between the two. An uncertainty of 2.1\% in the integrated luminosity affects the cross-section measurement and the MC-based estimate of backgrounds. It is determined following the same methodology as that detailed in Ref.~\cite{DAPR-2013-01} based on a calibration of the luminosity scale using $x$--$y$ beam-separation scans performed in August 2015. The beam energy uncertainty of 0.66\% (from Ref.~\cite{Wenninger:1546734}) gives a 1.7\% uncertainty in the theoretical cross section, which is not accounted for in the predictions quoted in this Letter.

Uncertainties in the $C$ factor due to theoretical sources are also included. The uncertainties associated with PDFs are taken as the largest of either the \cttennlo\ eigenvector uncertainty band at $68\%$ confidence level, or the difference among the central values of \cttennlo, \mstw~\cite{MSTW} and \nnpdf~\cite{Ball:2014uwa} PDFs. 
The uncertainty associated with higher-order QCD corrections is estimated by varying renormalisation ($\mu_{\rm R}$) and factorisation ($\mu_{\rm F}$) scales independently by factors of 2 and 0.5 with the constraint $0.5 \le \mu_{\rm F} / \mu_{\rm R} \le 2$. 
The effects of parton shower, hadronisation and underlying event models (referred to here as {\it parton shower} for simplicity) are accounted for by comparing the default MC prediction for $WW$ production, which uses \pythiaeight\ for modelling of these effects, with the prediction obtained with the models implemented in \herwigpp.

A full list of systematic uncertainties and their impact on the cross-section measurement is given in Table ~\ref{table:xsecSys}.


\section{The fiducial cross-section measurement}
\label{sec:result}

\begin{table}[t]
\begin{center}
\begin{tabular}{l*{3}{r@{$\,\pm\,$}l}}%
\hline\hline
Process & \multicolumn{2}{c}{Signal region} & \multicolumn{2}{c}{Top-quark} & \multicolumn{2}{c}{Drell--Yan}\\
~ & \multicolumn{2}{c}{~} & \multicolumn{2}{c}{control region} & \multicolumn{2}{c}{control region}\\
\hline\hline
$WW$ signal             & 997 & 69                         & 49  & 12                          & 75.3  & 5.4           \\
Drell--Yan               & 62  & 23                          & 49  & 29                          & 1568& 45                  \\
$t\bar{t}$+single top   & 177 & 33                          & 2057& 81                          & 3.5   & 1.6           \\
$W$+jets/multi-jet        & 78  & 41                          & 70  & 55                          & 0   & 17                  \\
Other dibosons          & 38  & 12                          & 6.3   & 3.5                       & 19.2 & 6.1           \\
\hline
Total                   & 1351 & 37                         & 2232 & 47                         & 1666  & 41 \\
\hline
Data & \multicolumn{2}{l}{{~~~~~~1351}}  & \multicolumn{2}{l}{{~~~~~~2232}}  & \multicolumn{2}{l}{{~~~~~1666}} \\
\hline\hline     \\
\end{tabular}
\vspace{0.1cm}
\end{center}
\vspace{-0.5cm}
\caption{ Observed number of events in data and estimated numbers of events from signal and background processes in the signal and control regions. The numbers of events from signal and background processes are the result of the simultaneous fit, i.e. are constrained to match the data in the signal and control regions. The quoted uncertainties account for statistical and systematic components on the number of events for each process and do not include the uncertainties in the $C$ factor. The correlations among processes for common systematic uncertainties are accounted for in the total uncertainties. 
} \label{tab:Postfit_EventsTable}
\end{table}

\begin{table}[t]
\centering
\begin{tabular}{ l | c   }
\hline
\hline
Sources of uncertainty & Relative uncertainty for $\sigma^{\rm fid}_{\WWemu}$ \\
\hline\hline
Jet selection and energy scale \& resolution           &  7.3\%  \\
$b$-tagging       &  1.3\%  \\
\met and \ptmiss         &  1.7\%  \\
Electron   &  1.0\%  \\
Muon    &  0.4\%  \\
\hline
Pile-up         &  0.9\%  \\
Luminosity      &  2.1\%  \\
\hline
Top-quark background theory       & 2.4\%  \\
Drell--Yan background theory       & 1.5\%  \\
$W$+jet and multi-jet background     & 3.8\%  \\
Other diboson backgrounds        & 1.1\% \\
\hline 
Parton shower 				&  3.1\%  \\
PDF 	 			&  0.2\%  \\
QCD scale 	 		&  0.2\%    \\
\hline 
MC statistics & 1.2\% \\
Data statistics  &  3.7\% \\ 
\hline \hline
Total uncertainty  			&  11\%	\\
\hline \hline
\end{tabular}
\caption{Breakdown of the relative uncertainties in the fiducial cross-section measurement as a result of the simultaneous fit to signal and control regions. ``Electron'' and ``Muon'' uncertainties include contributions from trigger, energy/momentum reconstruction, identification and isolation.}
\label{table:xsecSys}
\end{table}

The fiducial cross section $\sigma^{\mathrm{fid}}_{WW\rightarrow e\mu}$ is extracted by minimising a negative log-likelihood function, based on observed and expected numbers of events in the signal region, as defined by the signal event selection, and in the top-quark and Drell--Yan control regions, as defined in Section~\ref{sec:backgrounds}. 
The likelihood consists of a product of Poisson probability density functions for the orthogonal regions.  
This procedure allows a simultaneous measurement of the signal process cross section and of the contributions from the top-quark and Drell--Yan processes.
Systematic uncertainties are taken into account as constrained nuisance parameters in the log-likelihood function.  The methodology accounts for uncertainties and their correlations across signal and background processes.  
It is found that the Drell--Yan and top-quark processes need to be scaled relative to their MC predictions by $1.03 \pm 0.03$ and $0.875 \pm 0.035$ respectively to match the observed data yields in the corresponding control regions.
The uncertainties of the quoted scale factors are driven by the data statistics in the respective control regions and do not include modelling uncertainties on the respective processes.
The number of events observed in data and the estimated numbers of signal and background events together with their total uncertainties are reported in Table~\ref{tab:Postfit_EventsTable}. The correction factor $C$ is calculated to be $0.60\pm0.04$, where the uncertainty accounts for the systematic effects discussed in Section~\ref{sec:systematics}.  
The measured signal cross section is  

\begin{equation}
\sigma^{\rm fid}_{WW\rightarrow e\mu} = 529\pm 20 \stat \pm 50 \syst \pm 11 \lumi  {\rm ~fb}. \nonumber 
\label{EqResult}
\end{equation}

\begin{figure}[!t]
\begin{center}
\includegraphics[width=0.45\textwidth,keepaspectratio]{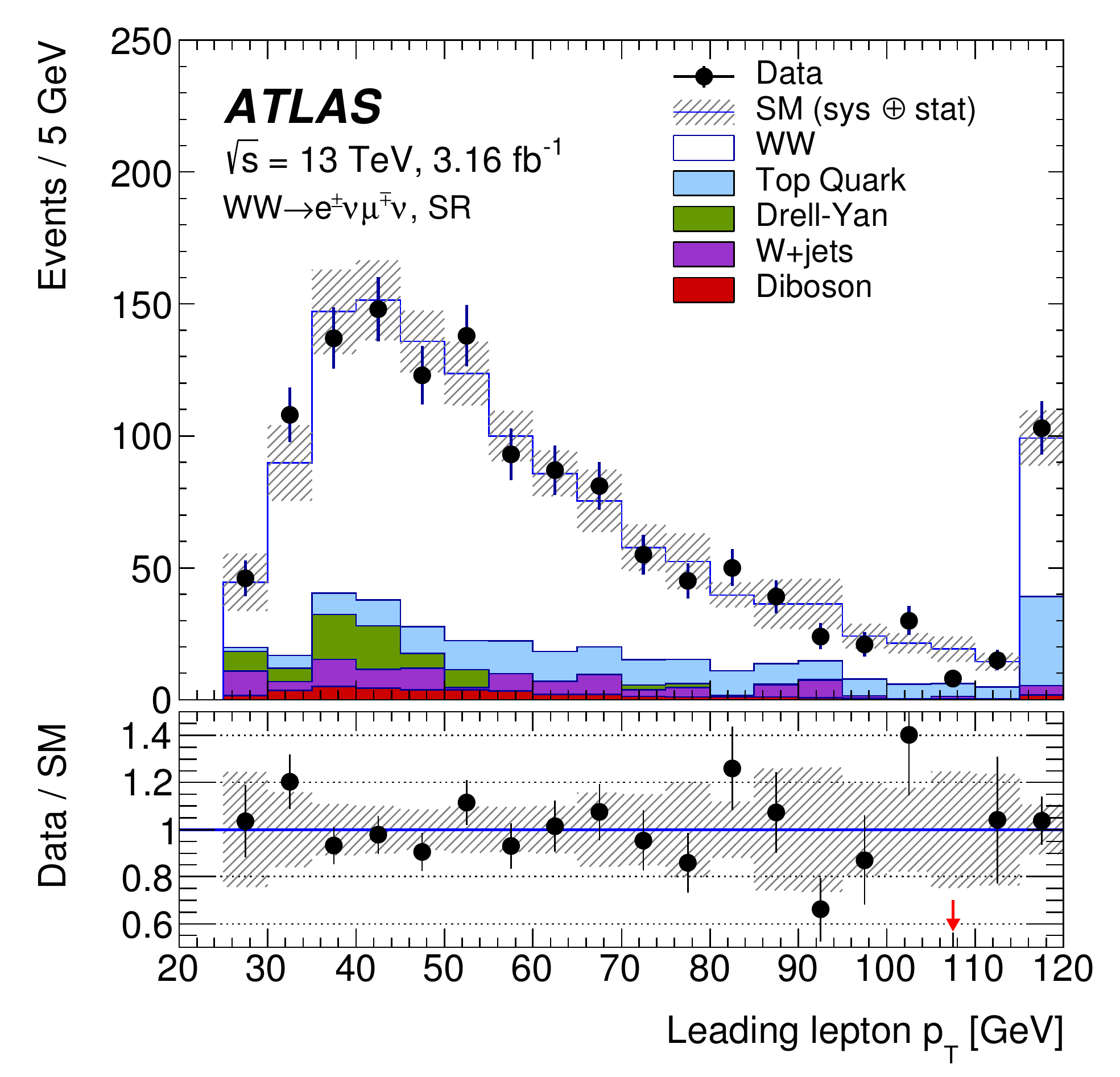}
\includegraphics[width=0.45\textwidth,keepaspectratio]{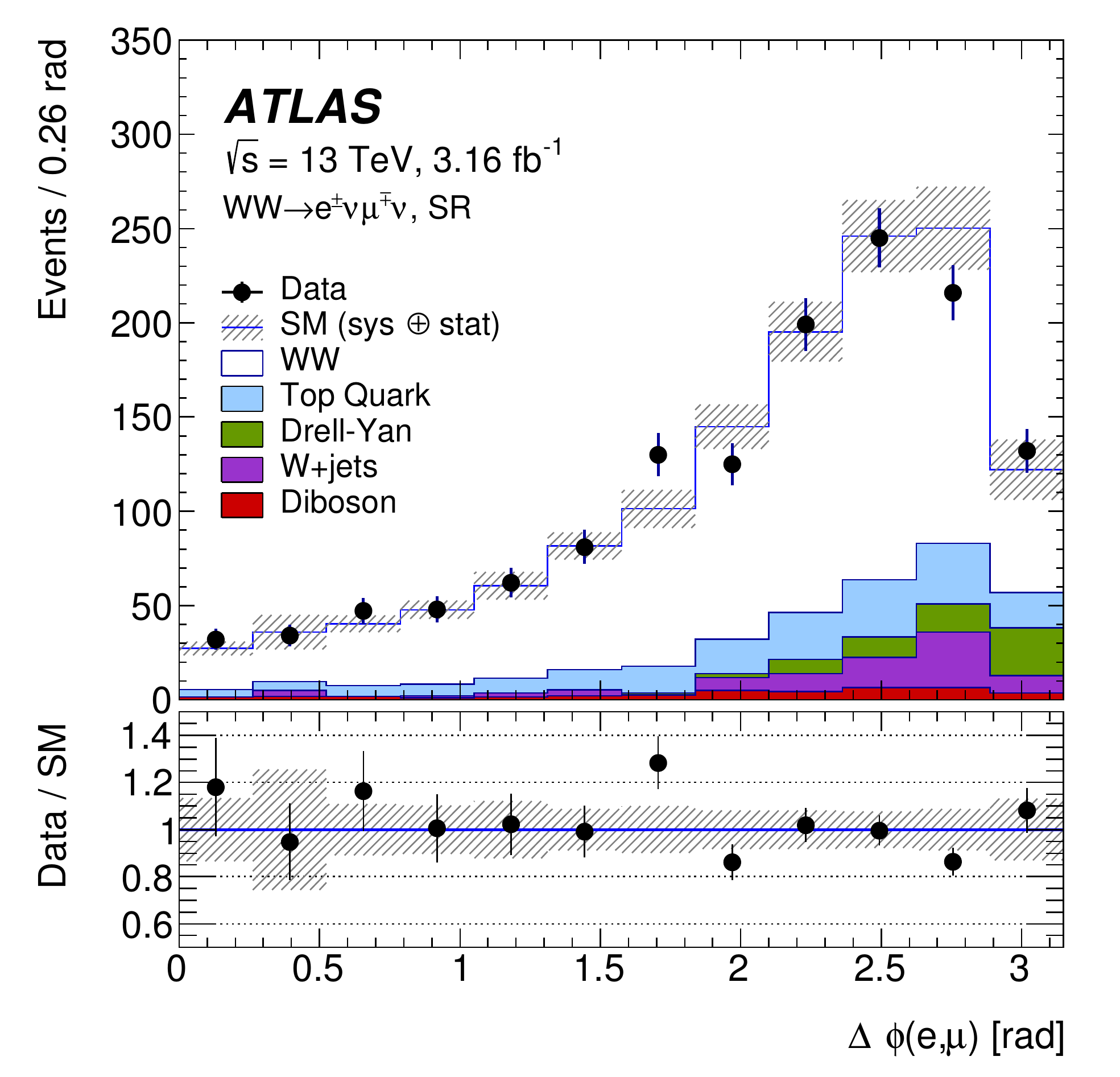}
\includegraphics[width=0.45\textwidth,keepaspectratio]{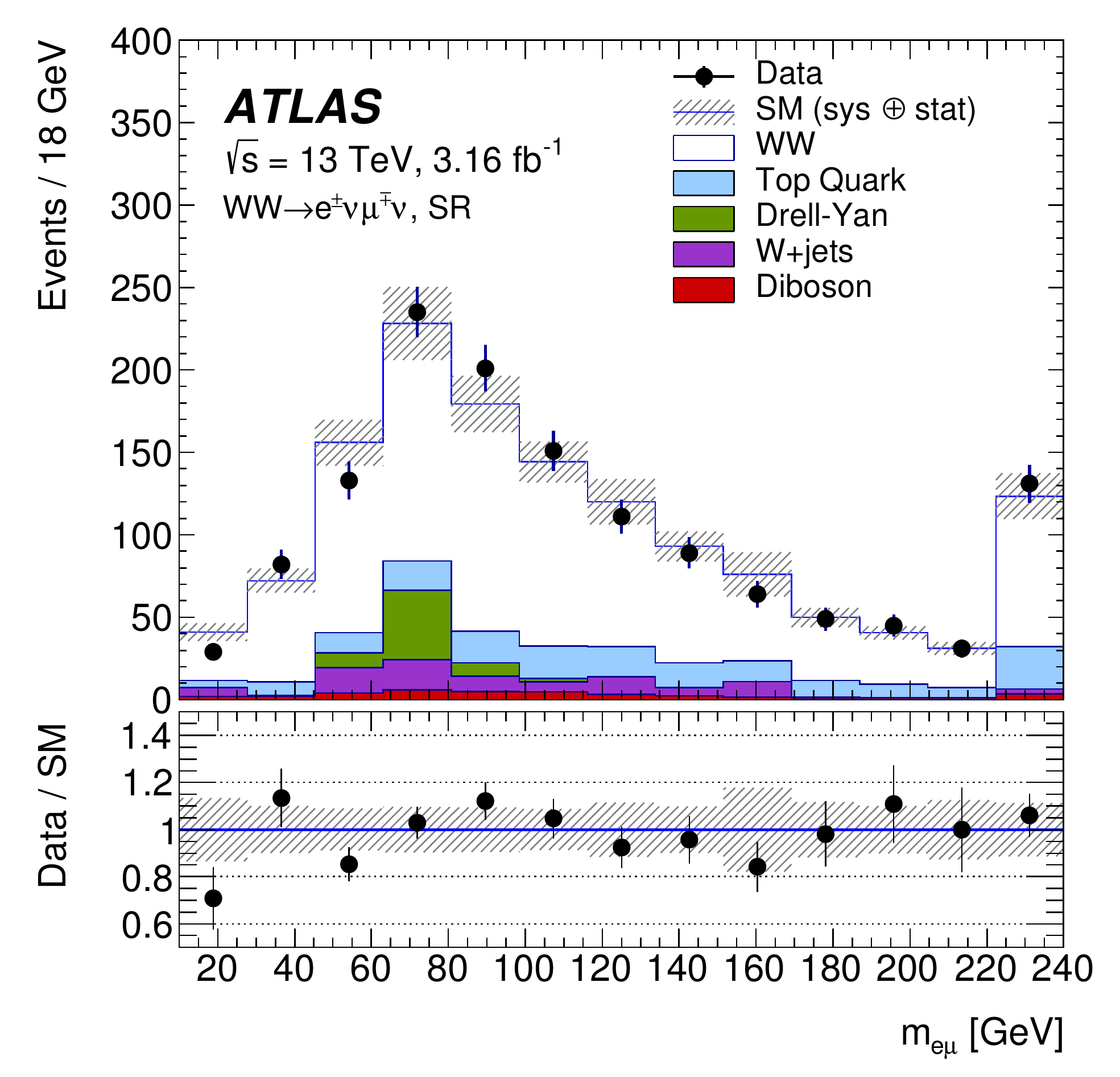}
\includegraphics[width=0.45\textwidth,keepaspectratio]{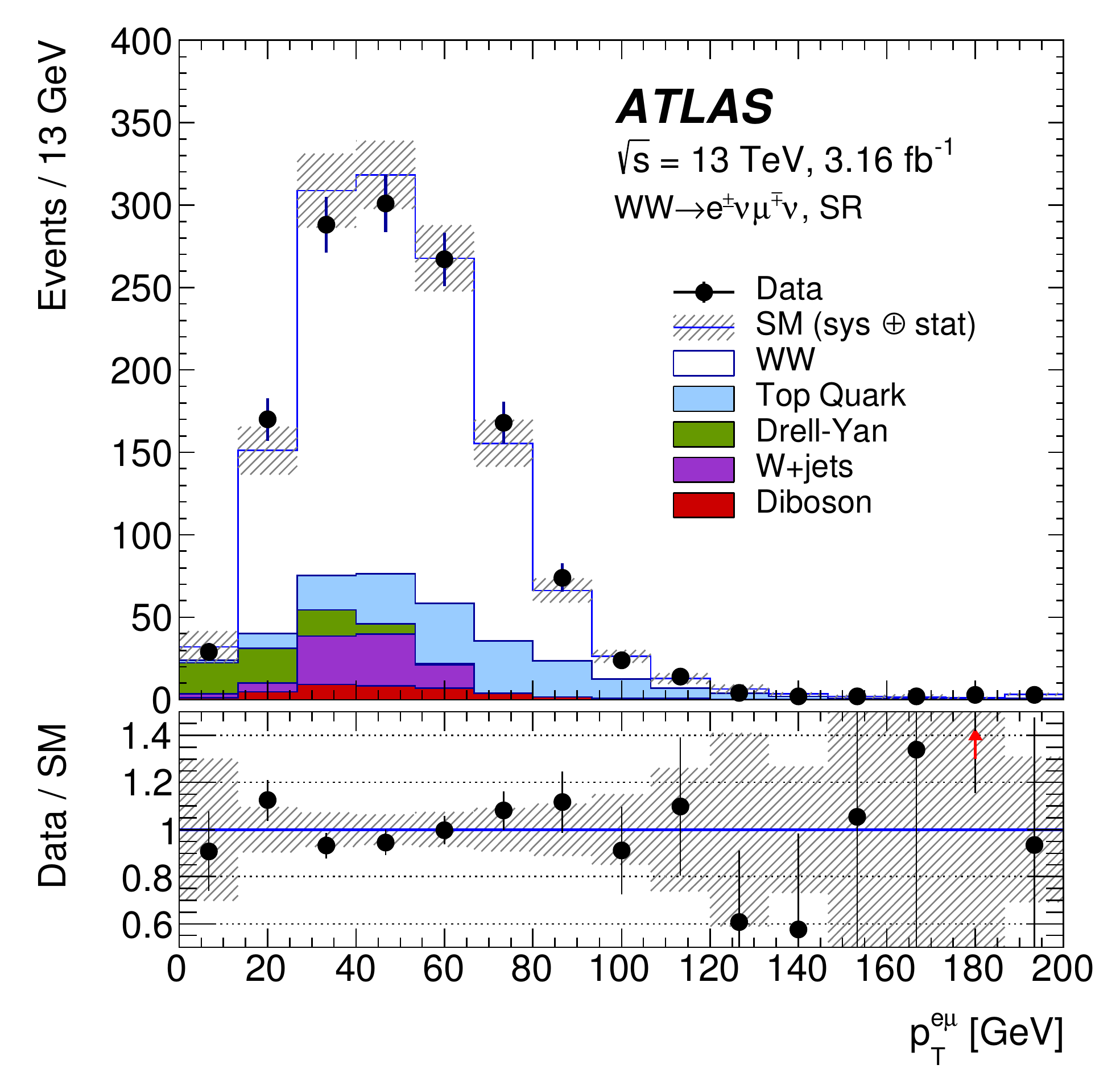}
\caption{ Lepton and dilepton kinematic variables in the signal region. In the figure $m_{e\mu}$ and $\pt^{e\mu}$ are the invariant mass and the transverse momentum of the $e$--$\mu$ system respectively, and $\Delta\phi(e,\mu)$ is the azimuthal angle between the two leptons. Data are shown together with the MC and data-driven predictions for the signal and background production processes after the fit to the data in the signal and control regions. The last bin in each distribution is the overflow. In the legend, SM stands for the total contribution of the estimated SM processes and the uncertainty band includes the MC statistical and systematic uncertainties as a result of the fit.
}
\label{fig:SR2post}
\end{center}
\end{figure}

The total uncertainty is dominated by systematic sources, as described in Section~\ref{sec:systematics}, of which the largest contribution originates from the experimental jet selection and calibration.  The correlations of the fit parameters in the signal and control regions are taken into account in the computation of the total uncertainties. The contributions to the relative uncertainty in the fiducial cross-section measurement are summarised in Table~\ref{table:xsecSys}. 
Figure~\ref{fig:SR2post} shows distributions of kinematic variables from data events in the signal region in comparison with the signal and background contributions estimated from the simultaneous fit to signal and control regions. 

\section{Theoretical predictions and ratio to the 8 \TeV\ measurement}
\label{sec:theoryComparison}
 
Theoretical predictions are calculated in the total phase space ($\sigma^{\mathrm{tot}}_{WW}$) and include the \qqtoWW, the non-resonant $gg\rightarrow WW$, and the resonant $gg\rightarrow H\rightarrow WW$ sub-processes. The \qqtoWW production cross section is known to $\mathcal{O}(\alpha_{\rm s}^2)$ (NNLO)~\cite{Gehrmann:2014fva,Grazzini:2016ctr}, the non-resonant $gg$ sub-process is known to $\mathcal{O}(\alpha_{\rm s}^3)$~\cite{Caola:2015rqy}, and the resonant $gg\rightarrow H \rightarrow WW$ cross section is calculated to $\mathcal{O}(\alpha_{\rm s}^5)$~\cite{Anastasiou:2016cez} taking into account the $H\rightarrow WW$ branching fraction~\cite{HiggsReport4}.
The sum of these sub-processes is denoted by nNNLO+H in the following. In its calculation, the interference between the three sub-processes is neglected. At the given orders of $\alpha_{\rm s}$ listed above, the \qqtoWW\ process does not interfere with either of the $gg$-induced processes and the interference between the $gg$-induced processes has little contribution to the cross section in the measured phase space. As in the 8 \TeV\ cross-section measurement, possible contributions from double parton interactions are not considered as their contribution is expected to be negligible \cite{STDM-2013-07}.

The renormalisation and factorisation scales are set to the $W$ boson mass for the $q\bar{q}$ and non-resonant $gg$ processes, and to $m_{H}/2$ for $gg\rightarrow H \rightarrow WW$. The uncertainties in the \qqtoWW cross section are estimated by varying the two scales independently by factors of 0.5 and 2 with the constraint $0.5 \le \mu_{\rm F} / \mu_{\rm R} \le 2$, while the uncertainties in the non-resonant and resonant $gg$ cross sections are estimated by simultaneously varying $\mu_{\rm R}$ and $\mu_{\rm F}$ by factors of 0.5 and 2. The uncertainties in $gg\rightarrow WW$ and $gg\rightarrow H \rightarrow WW$ processes include a $3.2\%$ contribution from PDF uncertainties computed in Ref.~\cite{Butterworth:2015oua}. For the \qqtoWW process, PDF uncertainties are estimated as the largest of either the \cttennlo\ eigenvector uncertainty band (at $68\%$ confidence level) or the difference among the central values of \cttennlo, \mstw~and \nnpdf~PDFs, amounting to $1.8\%$.
The uncertainties associated with the individual sub-processes are propagated to the $\sigma^{\mathrm{tot}}_{WW}$ prediction for the nNNLO+H combination: scale uncertainties of different processes are added linearly, while PDF uncertainties are considered uncorrelated across processes. 
The $q\bar{q}$ production makes up 87\% of the total cross section while the non-resonant and resonant $gg$ production sub-processes account for 5\% and 8\% respectively. 

\begin{table}[t]\renewcommand{\arraystretch}{1.3}
\begin{center}
\begin{tabular}{ p{4.1cm} | c | l | l | l}  
\hline
\hline
$pp\rightarrow WW$ sub-process &  Order of  &    \multicolumn{1}{c|}{$ \sigma^{\mathrm{tot}}_{WW} $}      & \multicolumn{1}{c|}{$A$}   & \multicolumn{1}{c}{$\sigma^{\mathrm{fid}}_{WW\rightarrow e\mu}$ }  \\  
 ~       & $\alpha_{\rm s}$  &    \multicolumn{1}{c|}{[pb] }              &  \multicolumn{1}{c|}{[\%]} & \multicolumn{1}{c}{[fb]}\\
\hline
\hline
$q\bar{q}$ ~\cite{Gehrmann:2014fva,Grazzini:2016ctr}  & $\mathcal{O}(\alpha_{\rm s}^2)$  &    111.1\phantom{0}$\pm$\phantom{0}2.8     & 16.20$\pm$0.13  &  422\phantom{0}\,$^{\,+\phantom{0}12}_{\,-\phantom{0}11}$ \\ 
\hline
$gg$ (non-resonant)  ~\cite{Caola:2015rqy}	& $\mathcal{O}(\alpha_{\rm s}^3)$		&   \phantom{0}\phantom{0}6.82\,$^{+\phantom{0}0.42}_{-\phantom{0}0.55}$ & 28.1\phantom{0}$^{+\phantom{0}2.7}_{-\phantom{0}2.3}$ &\phantom{0}44.9$\pm$7.2 \\
\hline
$gg\rightarrow H \rightarrow WW$~\cite{HiggsXS}\cite{Bagnaschi:2011tu} 	& $\mathcal{O}(\alpha_{\rm s}^{5})$ tot. / $\mathcal{O}(\alpha_{\rm s}^{3})$ fid.	 &  \phantom{0}10.45\,$^{+\phantom{0}0.61}_{-\phantom{0}0.79}$ & \phantom{0}4.5\phantom{0}$\pm$0.6  &  \phantom{0}11.0$\pm$2.1\\
\hline
\hline
$q\bar{q}$ + $gg$ (non-resonant) +  $gg\rightarrow H \rightarrow WW$			&nNNLO+H		&   128.4\phantom{0}\,$^{+\phantom{0}3.5}_{-\phantom{0}3.8}$ &  15.87$^{+\phantom{0}0.17}_{-\phantom{0}0.14}$ & 478\phantom{0}\,$\pm$17   \\
\hline
\hline
\end{tabular}
\caption{\label{tab:SigXsec} Theoretical predictions for the $WW$ cross-section sub-processes and their associated uncertainties in the full phase space ($\sigma^{\mathrm{tot}}_{WW}$) calculated up to the given order in $\alpha_{\rm s}$ together with the respective acceptance corrections ($A$) for the fiducial phase space and the fiducial cross sections ($\sigma^{\mathrm{fid}}_{WW\rightarrow e\mu}$). The resonant $gg\rightarrow H \rightarrow WW$ is calculated up to $\mathcal{O}(\alpha_{\rm s}^{5})$ for $\sigma^{\mathrm{tot}}_{WW}$ and to $\mathcal{O}(\alpha_{\rm s}^{3})$ for $\sigma^{\mathrm{fid}}_{WW\rightarrow e\mu}$ and $A$. A correction is applied to $\sigma^{\mathrm{fid}}_{WW\rightarrow e\mu}$ and $A$ to account for non-perturbative effects.  The quoted uncertainties include scale variations and PDF uncertainties, with the latter being evaluated at NLO. The scale uncertainties are treated as correlated, whereas PDF uncertainties are treated as uncorrelated between the $q\bar{q}$ and the $gg$-induced processes. The values of the branching ratio of leptonic $W$-boson decay used for each sub-process are those reported in their respective References, while for the resonant $gg$ sub-process $\mathcal{B}=0.1083$~\cite{alcaraz:2006mx} is used.
}
\end{center}
\end{table}

For direct comparison to the experimental result, theoretical predictions are also calculated in the same phase space as the measurement ($\sigma^{\mathrm{fid}}_{WW\rightarrow e\mu}$) for the $q\bar{q}$ and non-resonant $gg$ processes.
A correction of $0.972\pm 0.001$ is applied to parton-level calculations for $\sigma^{\mathrm{fid}}_{WW\rightarrow e\mu}$ to account for the contribution of non-perturbative effects due to multi-parton interactions and hadronisation. 
This correction was calculated by comparing the particle-level cross section as predicted by the MC simulation with one obtained with a dedicated event generation where these effects are disabled in \pythiaeight. 
The uncertainty includes the MC statistical uncertainty and the systematic component estimated by comparing the above correction with the one estimated with the non-perturbative model implemented in the \herwigpp\ MC event generator.
The calculations reported here do not include high-order electroweak corrections.
In Ref.~\cite{Biedermann:2016guo} it is estimated that electroweak corrections up to NLO reduce the $WW$ cross section by 3--4\% in a phase space close to the one used in this analysis.
The $q\bar{q}$ and non-resonant $gg$ fiducial cross sections are calculated with the programs presented in Refs.~\cite{Gehrmann:2014fva,Grazzini:2016ctr} and~\cite{Caola:2015rqy} respectively.
For the resonant $gg\rightarrow H \rightarrow WW$ process, no fiducial calculation is available at $\mathcal{O}(\alpha_{\rm s}^{5})$. Therefore, this fiducial cross section is calculated by correcting the cross section in the full phase space ($\sigma^{\mathrm{tot}}_{WW}$) by the geometrical and kinematic acceptance $A$ as determined using the MC event generator \powhegbox\ {\tt v2} interfaced to \pythiaeight\ for parton showering and non-perturbative effects and the branching ratio ($\mathcal{B}$) for fully leptonic final states, $\mathcal{B}=0.1083$~\cite{alcaraz:2006mx}:

\begin{equation}
  \sigma^{\mathrm{fid}}_{WW\rightarrow e\mu} = 2\times \sigma^{\mathrm{tot}}_{WW} \times {A} \times \mathcal{B}^{2}.
\label{xSectTot}
\end{equation}

In this determination of the $gg\rightarrow H \rightarrow WW$ fiducial cross section, uncertainties from PDFs and scale uncertainties are considered for both $A$ and $\sigma^{\mathrm{tot}}_{WW}$, while parton shower uncertainties are also estimated for $A$.
The $q\bar{q}$ and non-resonant $gg$ acceptances are calculated using the ratios of the respective fiducial cross section to the total cross section.  The uncertainties in the $A$ factors for $q\bar{q}$ and non-resonant $gg$ processes are estimated following the same methodology as for $\sigma^{\mathrm{tot}}_{WW}$ and considering both the scale and PDF uncertainties as correlated between $\sigma^{\mathrm{fid}}_{WW\rightarrow e\mu}$ and $\sigma^{\mathrm{tot}}_{WW}$. The total uncertainty in $A$ in the nNNLO+H calculation is then determined from the propagation of the $A$ factor uncertainties for the individual sub-processes.
The PDF uncertainties are found to be dominant and to lead to an uncertainty of $2.5\%$ and $3.2\%$ on $A$ for the $q\bar{q}$ and $gg$ sub-processes respectively.

\begin{figure}[t]
\begin{center}
\includegraphics[width=0.65\textwidth,keepaspectratio]{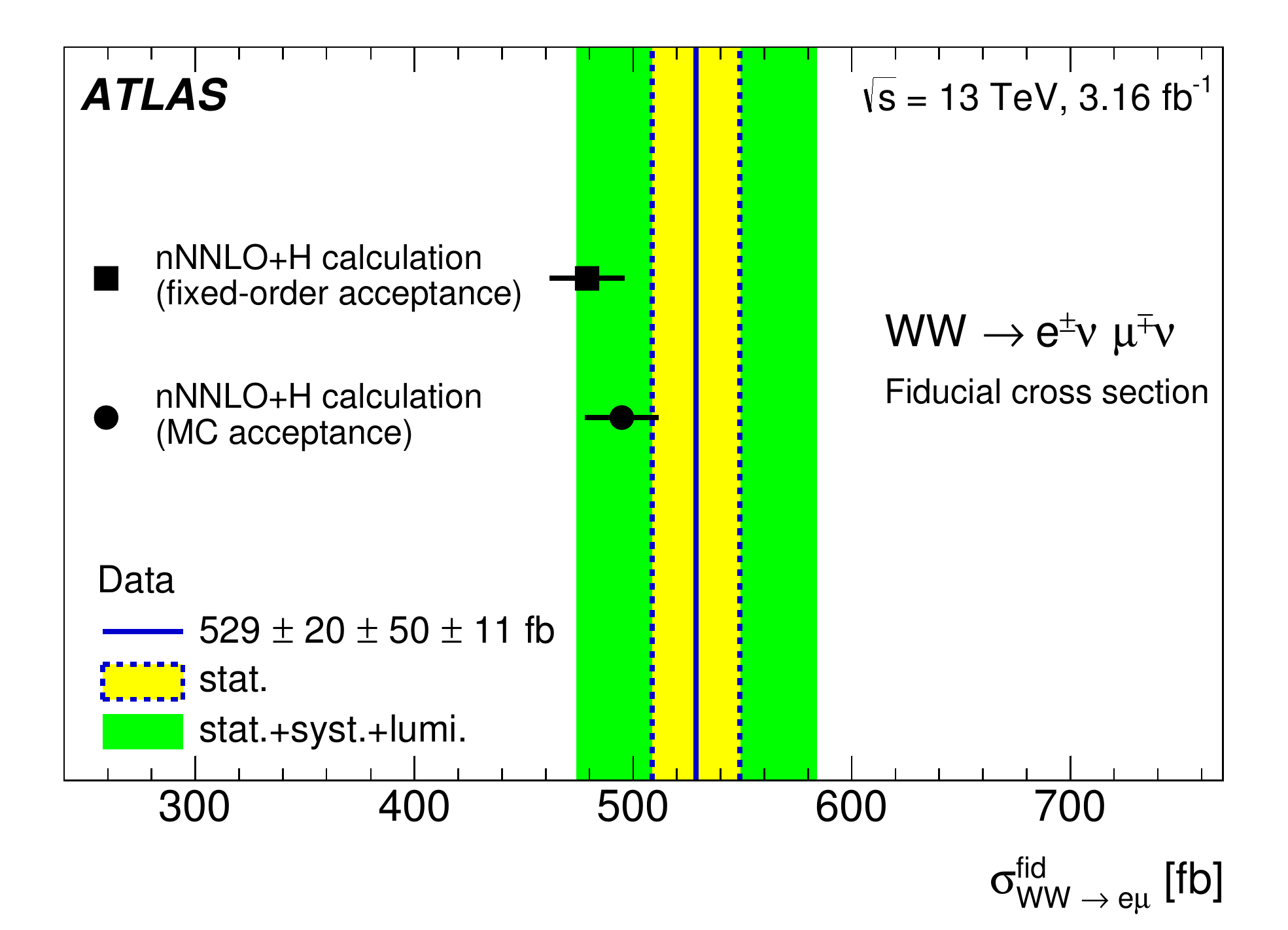}
\caption{ The measured fiducial cross section at $\sqrt{s}=$13 \TeV\ in comparison with the nNNLO+H prediction in the fiducial phase space with two different acceptance calculations. The vertical bands around the measurement indicate the statistical uncertainty (yellow) and the sum in quadrature of statistical, systematic and luminosity uncertainties (green). The beam energy uncertainty is not taken into account.}
\label{fig.FiducialCrossSections}
\end{center}
\end{figure}

The theoretical cross-section predictions for each production sub-process and the nNNLO+H combination in the total and fiducial phase spaces as well as the $A$ factors (corrected for non-perturbative effects) are given together with their estimated uncertainties in Table~\ref{tab:SigXsec}. Figure~\ref{fig.FiducialCrossSections} shows the comparison of the nNNLO+H prediction with the measurement presented in the previous section. Figure~\ref{fig.FiducialCrossSections} also reports, as an alternative prediction, the $\sigma^{\mathrm{tot}}_{WW}$ calculation for the nNNLO+H combination corrected by the acceptance $A$ calculated using the MC event generator \powhegbox\ {\tt v2} + \pythiaeight\ for the $q\bar{q}$ and resonant $gg\rightarrow H \rightarrow WW$ processes, and \sherpa\ {\tt v2.1.1} for the non-resonant $gg$ process.
In this calculation the acceptance factor is estimated to be $A$ = ($16.4 \pm  0.9$)~\% where the uncertainty includes the parton shower modelling (taken as the difference between \pythiaeight\ and \herwigpp\ showers), PDF uncertainty (estimated as the largest difference between the \cttennlo\ eigenvector uncertainty band and the \mstw~and \nnpdf~PDF central values), scale uncertainty associated with the jet veto requirement estimated as in Ref.~\cite{Stewart:2011cf} and the residual renormalisation and factorisation scale uncertainty (estimated by varying the two scales independently by factors of 2 and 0.5).

The nNNLO+H prediction agrees within uncertainties with the experimental cross-section measurement in the fiducial phase space.

The cross section in the full phase space ($\sigma^{\mathrm{tot}}_{WW}$) is determined by extrapolating the measurement in the fiducial phase space by inverting Eq.~(\ref{xSectTot}) and using the acceptance value from the nNNLO+H calculation as in Table~\ref{tab:SigXsec}: $\sigma^{\rm tot}_{WW} = 142\pm 5 \rm~(stat.) \pm 13 \syst \pm 3 \rm~(lumi.)  \rm ~pb$. This is in agreement with the nNNLO+H prediction of $128.4^{+3.5}_{-3.8} {\rm ~pb}$.

\begin{figure}[t]
\begin{center}
\includegraphics[width=0.65\textwidth,keepaspectratio]{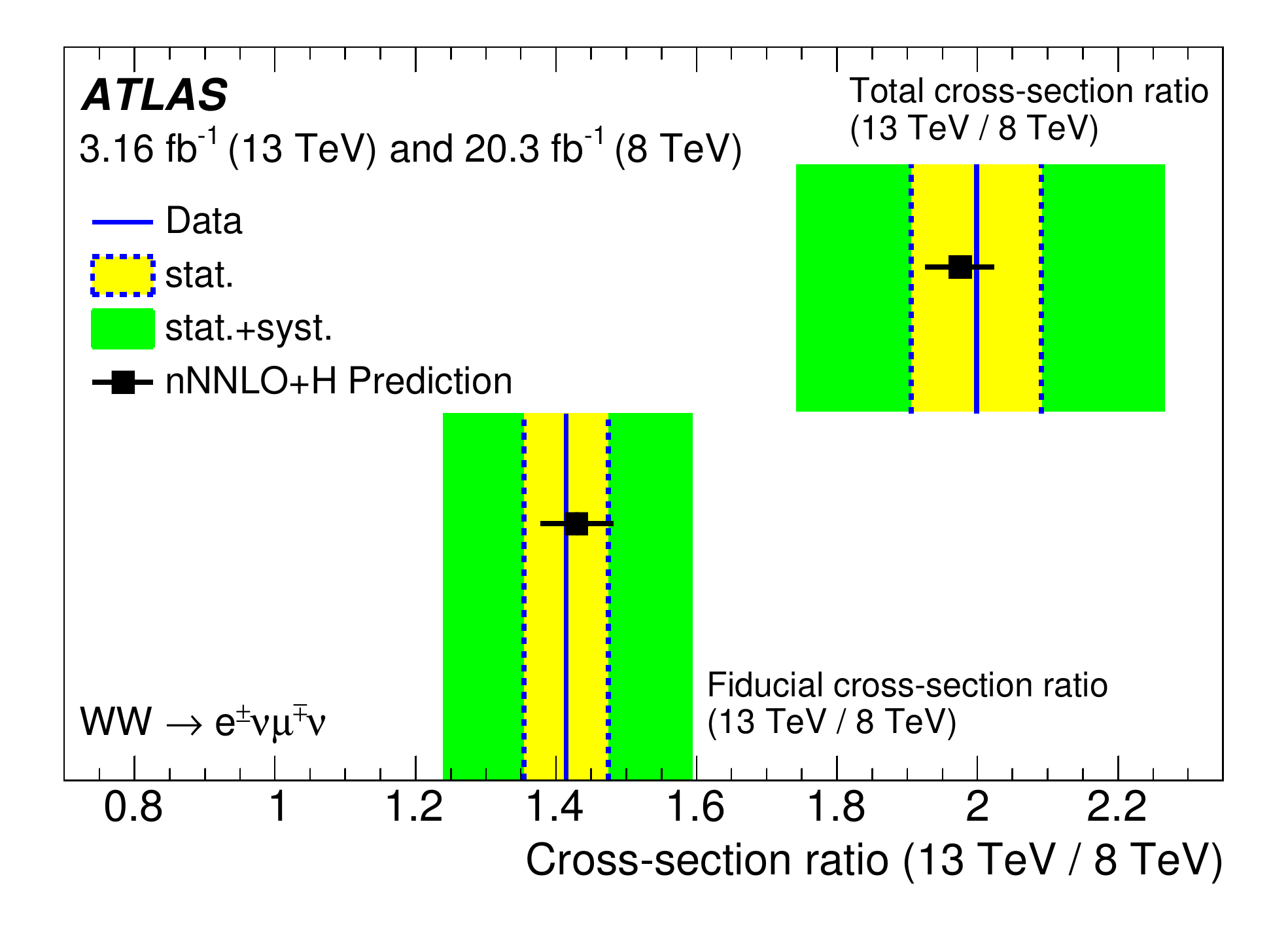}
\caption{Measurements of the ratios of cross sections at the two centre-of-mass energies of 13 and 8 \TeV\ in the fiducial and total phase spaces. For the 8 \TeV\ cross sections the results from Ref.~\cite{STDM-2013-07} are used. The measurements are compared to the nNNLO+H predictions for the ratios of cross sections in the fiducial phase spaces of the two analyses at 13 and 8 \TeV\ and in the total phase space, with their respective uncertainties. The beam energy uncertainty is not taken into account. }
\label{fig:DataTheoryFid}
\end{center}
\end{figure}

Using the fiducial cross section measured for $WW\rightarrow e\mu$ production at 8 \TeV\ centre-of-mass energy~\cite{STDM-2013-07}  in the fiducial phase space detailed in Ref.~\cite{STDM-2013-07}, the ratio of cross sections at the two centre-of-mass energies of 13 and 8 \TeV\ is:

\begin{equation}
\frac{ \sigma^{\mathrm{fid}}_{\mathrm{13~\TeV},{WW\rightarrow e\mu}}} { \sigma^{\mathrm{fid}}_{\mathrm{8~\TeV},{WW\rightarrow e\mu}}} = 1.41 \pm 0.06 \mathrm{\stat} \pm 0.16 \mathrm{\syst} \pm 0.04 \mathrm{(lumi.)}. \nonumber
\label{xSectRatio}
\end{equation}
All uncertainties are treated as uncorrelated between the measurements at the two beam energies: no attempt is made to exploit the jet energy scale correlations between the two data-taking periods at different beam energies. The same ratio is calculated for the total cross sections at 13 and 8 \TeV\ and is found to be $2.00 \pm 0.08 \mathrm{\stat}$ $^{+0.25}_{-0.24} \mathrm{(syst.)} \pm 0.06 \mathrm{(lumi.)} $.  Figure~\ref{fig:DataTheoryFid} shows the measured ratios of cross sections in the fiducial and total phase spaces and the comparison with their respective nNNLO+H predictions with scale uncertainties treated as correlated between the two centre-of-mass energies, while the PDF uncertainties are considered uncorrelated. The predictions for the ratio in the fiducial and total phase spaces are $1.43 \pm 0.05$ and $1.98 \pm 0.05$ respectively, and are in agreement with the experimental results.

\section{Conclusions}
\label{sec:conclusion}
The cross section for production of $W^+W^-$ pairs in $pp$ collisions at $\sqrt{s}=13$ \TeV\ is measured in a fiducial phase space of the $e\mu$ final state in which events with reconstructed jets are excluded. The data used in the analysis correspond to an integrated luminosity of 3.16 fb$^{-1}$ collected by the ATLAS detector at the LHC in 2015.
The measurement is made in a relatively pure signal region with the contamination from the dominant background processes estimated using data in dedicated control regions. 
The measured cross section is $ 529 \pm 20 \stat \pm 50 \syst \pm 11 \lumi$ fb and is found to be consistent with the most up-to-date SM predictions that include high-order QCD effects.  Furthermore, the ratio of the measured fiducial cross sections at 13 and 8 \TeV\ centre-of-mass energies is compared to the theory predictions with reduced uncertainties, thanks to their cancellation in the ratio.

\section*{Acknowledgements}


We thank CERN for the very successful operation of the LHC, as well as the
support staff from our institutions without whom ATLAS could not be
operated efficiently.

We acknowledge the support of ANPCyT, Argentina; YerPhI, Armenia; ARC, Australia; BMWFW and FWF, Austria; ANAS, Azerbaijan; SSTC, Belarus; CNPq and FAPESP, Brazil; NSERC, NRC and CFI, Canada; CERN; CONICYT, Chile; CAS, MOST and NSFC, China; COLCIENCIAS, Colombia; MSMT CR, MPO CR and VSC CR, Czech Republic; DNRF and DNSRC, Denmark; IN2P3-CNRS, CEA-DSM/IRFU, France; GNSF, Georgia; BMBF, HGF, and MPG, Germany; GSRT, Greece; RGC, Hong Kong SAR, China; ISF, I-CORE and Benoziyo Center, Israel; INFN, Italy; MEXT and JSPS, Japan; CNRST, Morocco; FOM and NWO, Netherlands; RCN, Norway; MNiSW and NCN, Poland; FCT, Portugal; MNE/IFA, Romania; MES of Russia and NRC KI, Russian Federation; JINR; MESTD, Serbia; MSSR, Slovakia; ARRS and MIZ\v{S}, Slovenia; DST/NRF, South Africa; MINECO, Spain; SRC and Wallenberg Foundation, Sweden; SERI, SNSF and Cantons of Bern and Geneva, Switzerland; MOST, Taiwan; TAEK, Turkey; STFC, United Kingdom; DOE and NSF, United States of America. In addition, individual groups and members have received support from BCKDF, the Canada Council, CANARIE, CRC, Compute Canada, FQRNT, and the Ontario Innovation Trust, Canada; EPLANET, ERC, ERDF, FP7, Horizon 2020 and Marie Sk{\l}odowska-Curie Actions, European Union; Investissements d'Avenir Labex and Idex, ANR, R{\'e}gion Auvergne and Fondation Partager le Savoir, France; DFG and AvH Foundation, Germany; Herakleitos, Thales and Aristeia programmes co-financed by EU-ESF and the Greek NSRF; BSF, GIF and Minerva, Israel; BRF, Norway; CERCA Programme Generalitat de Catalunya, Generalitat Valenciana, Spain; the Royal Society and Leverhulme Trust, United Kingdom.

The crucial computing support from all WLCG partners is acknowledged gratefully, in particular from CERN, the ATLAS Tier-1 facilities at TRIUMF (Canada), NDGF (Denmark, Norway, Sweden), CC-IN2P3 (France), KIT/GridKA (Germany), INFN-CNAF (Italy), NL-T1 (Netherlands), PIC (Spain), ASGC (Taiwan), RAL (UK) and BNL (USA), the Tier-2 facilities worldwide and large non-WLCG resource providers. Major contributors of computing resources are listed in Ref.~\cite{ATL-GEN-PUB-2016-002}.

\printbibliography

\newpage 
\begin{flushleft}
{\Large The ATLAS Collaboration}

\bigskip

M.~Aaboud$^\textrm{\scriptsize 137d}$,
G.~Aad$^\textrm{\scriptsize 88}$,
B.~Abbott$^\textrm{\scriptsize 115}$,
J.~Abdallah$^\textrm{\scriptsize 8}$,
O.~Abdinov$^\textrm{\scriptsize 12}$,
B.~Abeloos$^\textrm{\scriptsize 119}$,
R.~Aben$^\textrm{\scriptsize 109}$,
O.S.~AbouZeid$^\textrm{\scriptsize 139}$,
N.L.~Abraham$^\textrm{\scriptsize 151}$,
H.~Abramowicz$^\textrm{\scriptsize 155}$
% ...
\bigskip
\\
% ...
$^{8}$ Department of Physics, The University of Texas at Arlington, Arlington TX, United States of America\\
% ...
$^{12}$ Institute of Physics, Azerbaijan Academy of Sciences, Baku, Azerbaijan\\
% ...
$^{88}$ CPPM, Aix-Marseille Universit{\'e} and CNRS/IN2P3, Marseille, France\\
% ...
$^{109}$ Nikhef National Institute for Subatomic Physics and University of Amsterdam, Amsterdam, Netherlands\\
% ...
$^{115}$ Homer L. Dodge Department of Physics and Astronomy, University of Oklahoma, Norman OK, United States of America\\
% ...
$^{119}$ LAL, Univ. Paris-Sud, CNRS/IN2P3, Universit{\'e} Paris-Saclay, Orsay, France\\
% ...
$^{137}$ $^{(a)}$ Facult{\'e} des Sciences Ain Chock, R{\'e}seau Universitaire de Physique des Hautes Energies - Universit{\'e} Hassan II, Casablanca; $^{(b)}$ Centre National de l'Energie des Sciences Techniques Nucleaires, Rabat; $^{(c)}$ Facult{\'e} des Sciences Semlalia, Universit{\'e} Cadi Ayyad, LPHEA-Marrakech; $^{(d)}$ Facult{\'e} des Sciences, Universit{\'e} Mohamed Premier and LPTPM, Oujda; $^{(e)}$ Facult{\'e} des sciences, Universit{\'e} Mohammed V, Rabat, Morocco\\
% ...
$^{139}$ Santa Cruz Institute for Particle Physics, University of California Santa Cruz, Santa Cruz CA, United States of America\\
% ...
$^{151}$ Department of Physics and Astronomy, University of Sussex, Brighton, United Kingdom\\
% ...
$^{155}$ Raymond and Beverly Sackler School of Physics and Astronomy, Tel Aviv University, Tel Aviv, Israel\\
% ...
$^{a}$ Also at Department of Physics, King's College London, London, United Kingdom\\
$^{b}$ Also at Institute of Physics, Azerbaijan Academy of Sciences, Baku, Azerbaijan\\
$^{c}$ Also at Novosibirsk State University, Novosibirsk, Russia\\
$^{d}$ Also at TRIUMF, Vancouver BC, Canada\\
$^{e}$ Also at Department of Physics {\&} Astronomy, University of Louisville, Louisville, KY, United States of America\\
% ...
\end{flushleft}



\end{document}